\documentclass[a4paper]{jpconf}
\usepackage{graphicx}
\usepackage{amsmath}
\usepackage{amssymb}

\newcommand{\Ran}{{\rm Ran}}
\newcommand{\diag}{{\rm diag}}
\newcommand{\Id}{{\rm Id}}

\newcommand{\Const}{{\rm Const.}}

\begin{document}
\title{Generalized Pareto optimum 
and semi-classical spinors}

\author{M~Rouleux}

\address{$^2$ Aix-Marseille University, Universit\'e de Toulon, CNRS, CPT, Marseille, France}

\ead{rouleux@univ-tln.fr}

\begin{abstract}
In 1971, S.Smale presented a generalization of Pareto optimum 
he called the critical Pareto set. The underlying motivation was to 
extend Morse theory to several functions, i.e. to find a Morse theory for $m$ differentiable functions
defined on a manifold $M$ of dimension $\ell$. We use this framework to 
take a $2\times2$ Hamiltonian ${\cal H}={\cal H}(p)\in C^\infty(T^*{\bf R}^2)$ to its normal form near a singular point
of the Fresnel surface. Namely we say that ${\cal H}$ has the Pareto property if 
it decomposes, locally, up to a conjugation with regular matrices, 
as ${\cal H}(p)=u'(p)C(p)(u'(p))^*$, where $u:{\bf R}^2\to{\bf R}^2$ has 
singularities of codimension 1 or 2, and $C(p)$ is a regular Hermitian matrix
(``integrating factor''). In particular this applies in certain cases to the matrix Hamiltonian
of Elasticity theory and its (relative) perturbations of order 3 in momentum at the origin.
\end{abstract}

\section{\bf Matrix valued Hamiltonian systems and crossing of modes} 
\medskip
Here we recall some well known facts about generic normal forms for a matrix Hamiltonian near a crossing of modes.
Consider a real valued symmetric Hamiltonian 
${\cal H}\in C^\infty(T^*{\bf R}^d)\otimes{\bf R}^n$.
We may also replace ${\cal H}(x,\xi)$ by a more general symbol (in a suitable class) with asymptotic sum
${\cal H}(x,\xi;h)={\cal H}_0(x,\xi)+h{\cal H}_1(x,\xi)+\cdots$, 
and consider its semi-classical Weyl quantization
$${\cal H}^w(x,hD_x;h)U(x;h)=(2\pi h)^{-d}\int\int e^{i(x-y)\eta/h} {\cal H}({x+y\over2},\eta;h)U(y)\,dy\,d\eta$$
This defines an operator essentially self-adjoint on the space of square-integrable ``spinors'' $L^2({\bf R}^d)\otimes{\bf C}^n$.
For simplicity we keep denoting by ${\cal H}$ the principal symbol ${\cal H}_0$ which actually
plays the main role in the present analysis. 

Let $\Delta(x,\xi)=\det {\cal H}(x,\xi)$, and $N=\{(x,\xi)\in T^*{\bf R}^d, \Delta(x,\xi)=0\}$
be the characteristic variety. 

At points of $N$, the polarisation space 
$\ker H(x,\xi)$ has positive dimension $k\geq1$. 

For $k=1$, $\{(x,\xi)\in N: d\Delta\neq 0\}$ is a smooth hypersurface, and modulo an elliptic 
factor, we can reduce ${\cal H}$ to a scalar symbol with simple
characteristics; the corresponding operator, of real principal type, can again be locally conjugated to $hD_{x_1}$ by an
elliptic Fourier integral operator (FIO); this induces well-known polarization properties for solutions of the original system.

Effects which are truly specific for systems occur therefore when $k\geq2$; let
$\Sigma=\{(x,\xi)\in N: d\Delta= 0\}$ be the singular part of $N$. 

For generic Hamiltonian ${\cal H}$,
$\Sigma$ is a stratified set.  Splitting off elliptic summands, $H$ can be (generically) reduced to a $2\times2$ system.
For a $2\times2$ symmetric system $\begin{pmatrix}a+f&b\\b&-a+f\end{pmatrix}$, one should generically move 2 parameters 
to bring it to a matrix with an eigenvalue of multiplicity 2, and one more to bring this eigenvalue to 0; thus the 
top (larger) stratum of $\Sigma$ is of codim 3. On directions transverse to $\Sigma$,
$N$ looks like a quadratic cone. In the next item we review some known results, making these observations more precise.\\

\noindent {\it a) Classical and semi-classical normal forms for generic $H$}\\

We first mention Arnold's \cite{Ar1} normal forms for (generic) $\Delta=\det{\cal H}$ near the stratum of $\Sigma$ of codim 3: at a conical point (i.e.
where $a=b=c=0$), there are symplectic coordinates $(p,q)$ such that 
modulo an elliptic factor, we can bring $\Delta$ to one of the following expressions~: 
\begin{eqnarray*} 
&\Delta(x,\xi)\equiv p_1q_1-p_2^2 \quad  \hbox{hyperbolic conical point}\\
&\Delta(x,\xi)\equiv p_1^2+q_1^2-p_2^2 \quad \hbox{elliptic conical point}
\end{eqnarray*}
When $d\geq 3$  Braam and Duistermaat \cite{BrDu1} \cite{BrDu2} have further brought ${\cal H}$ to its normal form
near the stratum of $\Sigma$ of codim 3
at a conical point~:
Modulo elliptic summands, we can reduce to the case where ${\cal H}$ is a $2\times2$ real symmetric matrix, and 
in the sense of quadratic forms, i.e. by replacing ${\cal H}$ by ${\cal A}^*{\cal H}{\cal A}$,
where ${\cal A}$ is elliptic, there are symplectic coordinates $(p,q)$ such that
\begin{eqnarray*}
&{\cal A}^*{\cal H}{\cal A}=\begin{pmatrix}-p_1+p_2&q_1p_3\\ q_1p_3&p_1+p_2\end{pmatrix} \  \hbox{hyperbolic conical point}\\
&{\cal A}^*{\cal H}{\cal A}=\begin{pmatrix}p_1-p_2&q_2p_3\\ q_2p_3&p_1+p_2\end{pmatrix} \  \hbox{elliptic conical point}
\end{eqnarray*}

Moreover these normal forms extend naturally to quantization of Weyl symbols, composing ${\cal A}$ by a 
suitable FIO ${\cal U}$ microlocally defined near the conical point, and associated with the canonical
change of coordinates $(x,\xi)\mapsto(q,p)$. Namely, in the hyperbolic case
$${\cal U}^*({\cal A}^*{\cal H}{\cal A})^w{\cal U}=\begin{pmatrix}-hD_{q_1}+hD_{q_2}&q_1hD_{q_3}\\ q_1hD_{q_3}&hD_{q_1}+hD_{q_2}\end{pmatrix}$$
and similarly in the elliptic case. 

Normal forms are useful for studying the Hamiltonian flow of ${\cal H}(x,\xi)$ or finding quasi-modes
for ${\cal H}(x,hD_x)$ in terms of classical functions.
These normal forms are structurally stable, i.e. they are not affected by small perturbations of a generic ${\cal H}$ in the
$C^\infty(T^*{\bf R}^d)\otimes{\bf R}^n$ topology. Even more precise results are available in 1-D, see e.g. \cite{Ro}
in the framework of Born-Oppenheimer approximation, or \cite{DuGy}, \cite{BenMhRo} for Bogoliubov-de Gennes Hamiltonian.\\

\noindent {\it b) Particular cases: $k=2$, codim 3 singularities}\\

In many physical situations however, the genericity assumption is not fulfilled, and special reductions should be carried out~:\\

\noindent $\bullet$ {\it Conical refraction in 3-D} Recall from \cite{Iv} in dimension $d=3$ that the normal form is given by the symmetric matrix
\begin{equation}
{\cal H}=\begin{pmatrix}p_1+p_2&p_3\\ p_3&p_1-p_2\end{pmatrix}
\end{equation}
which is independent of space variables. This is not structurally stable, because the singular part $\Sigma$ of $N$ is involutive.\\

\noindent $\bullet$ {\it Graphene Hamiltonian in 2-D}. The Hamiltonian is a complex Hermitian matrix \cite{FeWe}:
\begin{equation}
{\cal H}(p)=\begin{pmatrix}0&f(p_1,p_2)\\ \overline f(p_1,p_2)&0\end{pmatrix}=|\lambda(p)|\begin{pmatrix} 0&e^{-i\theta(p)}\\e^{-i\theta(p)}\end{pmatrix}
\end{equation}
Here $p=(p_1,p_2)\in {\bf R}^2$ are quasi-momenta, and the eigenvalues are explicitely given by
$$\lambda(p)=\pm t\sqrt{3+2\cos(\sqrt3 p_1a)+4\cos(\sqrt3 p_2a/2)\cos(3p_2a/2)}$$
Energy vanishes at Dirac points (2 Dirac points per hexagonal cell). The linearization at the Dirac point is of the form
\begin{equation}
{\cal H}(p)=v\begin{pmatrix}0&k_1-ik_2\\ k_1+ik_2&0\end{pmatrix}, \quad v=3t/(2a)
\end{equation}
\\

\noindent {\it c) Particular case: higher order singularities} \\

Here we are concerned in the case where $\Delta$ vanishes of order 4 at the conical point (leaving open the case $k=3)$. 
The physical example is the
Hamiltonian quadratic in the momenta from Elasticity Theory \cite{Bo2} in (2+1)-D, that is obtained in the following way. 

On the set of maps 
$\Gamma: {\bf R}_t\times{\bf R}_x^2\to{\bf R}^2, (t,x,y)\mapsto(\phi(x,y,t),\psi(x,y,t))$ with Sobolev regularity $H^1({\bf R}^3)$, we consider the
Lagrangian density:
\begin{equation*}
{\cal L}({x\choose y},{\phi_t\choose\psi_t},{\phi_x\choose\psi_x},{\phi_y\choose\psi_y})=
T-V={1\over2}(\phi_t^2+\psi_t^2)-{1\over2}(\phi_x^2+\psi_y^2)-{1\over2}A(x,y)(\phi_y+\psi_x)^2
\end{equation*}

Euler-Lagrange equations from the variational principle 
$\delta\int {\cal L}\,dt\,dx\,dy=0$ lead to extremal curves (rays) in the $(t,x,y)$-space, 
and the set of points $(t,x,y)$ connected to the origin in ${\bf R}^3$ by such a ray is called the ``world front''. 
A matter of interest are the singularities of the world front.

Applying Fourier transformation with respect to $(x,y,t)$ 
$$\phi\mapsto \widehat \phi, \ \phi_x\mapsto \xi\widehat \phi, \ \phi_y\mapsto \eta\widehat \phi, \ \phi_t\mapsto \tau\widehat \phi, 
\ \hbox{etc\dots}$$
switches from Lagrangian formulation to Hamiltonian formulation and leads 
to a Pseudo-differential Hamiltonian system (as a quadratic form in $(\widehat \phi,\widehat \psi)$) with principal symbol
\begin{equation}\label{1}
{\cal H}=\begin{pmatrix}\tau^2&0\\ 0&\tau^2\end{pmatrix}-
\begin{pmatrix}\xi^2+A(x,y)\eta^2&A(x,y)\xi\eta\\ A(x,y)\xi\eta&A(x,y)\xi^2+\eta^2\end{pmatrix}
\end{equation}
Genericity properties for this Hamiltonian imply as above that the singular set $\Sigma$ is of codimension 3, 
so that it can be brought to one of Arnold-Braam-Duistermaat normal forms (in 2-D) near $\Sigma$. 
However, genericity breaks down in the case of constant coefficient $A(x,y)=\Const $, which justify a direct approach.

In particular, for $A={1\over2}$, the spatial part has determinant 
$$\Delta(\xi,\eta)={1\over2}(\xi^2+\eta^2)^2$$
which vanishes of order 4 at $(\xi,\eta)=0$. 

Our purpose, precisely in this case, is to provide a normal form for ${\cal H}$ near $\tau=\xi=\eta=0$. 
This could be achieved by a straightforward diagonalisation of ${\cal H}$,
but our method carries naturally to (relative) perturbations
of this Hamiltonian, 
depending on $(\xi,\eta)$ alone; more naively this example is intended as an 
illustration of the role played by the generalized Pareto set.
 

\section{\bf Generalized Pareto optimum}
\medskip
A central problem in Differential Calculus consists in maximizing a function:
Morse theory on a smooth manifold provides a globalization of this problem.

Economists are rather concerned in maximizing simultenaously several ``utility functions'',
obtaining this way the notion of Pareto optimum in a free exchange economy.

In 1971, S.Smale presented a generalization of Pareto optimum 
he called the {\it critical Pareto set} \cite{Sm}. 
The main mathematical motivation is to find a Morse theory for $m$ differentiable functions
defined on a manifold $M$ of dimension $\ell$. Note that this problem is distinct from this of relative extrema
of a single function, 
which is relevant to Lagrange multipliers.\\

\noindent {\it a) ``Classical'' Pareto optimum in a free exchange economy}\\

Consider a free exchange economy consisting in $m$ consumers $i=1,\cdots,m$, 
and for each $i$, let $x_i=(x_i^{(1)},\cdots,x_i^{(J)})$ represent the (positive) amount of
goods $j=1,\cdots,J$, with $x_i^{(j)}\geq0$. We define the ``commodity space'' as 
$P=\{x=(x^{(1)},\cdots,x^{(J)})\in{\bf R}^J; x^{(j)}\geq0\}$. An unrestricted state of the economy is 
a point $\vec x=(x_1,\cdots,x_m)\in P^{m}$, but we may restrict to the affine space $M=\{\vec x\in P^{m}: \sum_{i=1}^mx_i^{(j)}=C^{(j)},
j=1,\cdots,J\}$
with total ressource $C^{(j)}$ of good $j$. 

Each consumer is supposed to have an utility function $u_i: P\to{\bf R}$ (generally an homogenous function of
$x^{(1)},\cdots,x^{(J)}$). Thus consumer $i$ prefers $x_i$ to $x'_i$ iff $u_i(x_i)>u_i(x'_i)$. The level sets 
$u_i^{-1}(c)$, $c>0$ are called ``indifference surfaces''. 

One considers exchanges in $M$ which will increase each $u_i$ on $M$.
A state $\vec x\in M$ is called ``Pareto optimal'' iff there is no $\vec x'\in M$ such that $u_i(x_i)\geq u_i(x'_i)$ for all $i$,
and $u_j(x_j)>u_j(x'_j)$ for some $j$. If $\vec x\in M$ is not Pareto, $\vec x$ is not economically stable.
For $m=1$ Pareto optimum equals the usual notion of a (constrained) maximum of $u:M\to{\bf R}, x\mapsto u(x)$. 

Physicists would replace everywhere the words ``maximizing'' by ``minimizing'', and ``total resource'' by ``total energy''.\\

\noindent {\it b) Generalized Pareto optimum in the sense of Smale.}\\

Here we do not only consider (joined) maxima, but also (joined) critical points. 
 
Let $M$ be a smooth manifold, dim $M=\ell$, and $u:M\to{\bf R}^m$ be $m$ smooth functions
defined by $u=(u_1,\cdots,u_m)$ (vector of ``utilities''). 

Let $H(x)=(u'(x))^{-1}({\bf R}_+^m)$, where
$u'(x):T_xM\to{\bf R}_+^m$ denotes the derivative (Jacobian) of $u$ at $x\in M$, and ${\bf R}_+^m$ the set of 
$v=(v_1,\cdots,v_m)\in{\bf R}^m$ with $v_j>0$ all $j$.\\

\noindent {\bf Definition 1} \cite{Sm}: We call $\Theta=\{x\in M: H(x)=\emptyset\}$ the Pareto critical set.\\

Thus the relation $x\in\Theta$  means that there is no smooth curve $\gamma:]-\epsilon,\epsilon[\to M$ starting at $x$, 
and such that $t\mapsto u_i\circ\gamma(t)$ increases for all $i$'s (gradient flow dynamics). 
For $m=1$, Pareto critical set is just
the set of critical points of $u$. 

If $u$ is a (single) Morse function on $M$, 
$\Theta$ is a discrete set, and the Hessian $u''(x)$ is defined intrinsically on $\Theta$. 
For $m>1$, $\Theta$ need not be discrete; but when it consists, as is usual, of a submanifold of dim $m-1$,
then $u''(x)$ is still intrinsically defined on $\Theta$ and valued in the 1-D space ${\bf R}^m/\Ran u'(x)$. 

The open subset $\Theta_S\subset\Theta$
of stable points (classical Pareto set), which reduces for $m=1$ to the set of local maxima of $u$, plays a special r\^ole. \\

\noindent $\bullet$ {\it Paradigm of Pareto critical in 2-D}: the immersed Klein bottle in ${\bf R}^3$.\\

The paradigm of a Morse function on a compact 2-manifold is the ``height function'' on the embedded torus,
and its (Pareto) critical points are the minimum, 2 saddles, and maximum. Similarly Pareto critical set for 2 
functions of 2 variables will be obtained from Klein bottle [Wan], by projecting the immersed bottle in ${\bf R}^3$ onto
a suitable plane, and looking at the ``joined extrema'' of the coordinates functions on the projection plane.
Another, more convenient immersion of Klein bottle in ${\bf R}^3$ (though not so easy to visualize)
is the so called ``figure eight'' or ``bagel'' immersion, given by 
\begin{eqnarray*}
&x_1=(r+\cos{\theta\over2}\sin v-\sin{\theta\over2}\sin 2v)\cos\theta\\
&x_2=(r+\cos{\theta\over2}\sin v-\sin{\theta\over2}\sin 2v)\sin\theta\\
&x_3=\sin{\theta\over2}\sin v+\cos{\theta\over2}\sin 2v
\end{eqnarray*}
with $r>2$ a parameter, $-\pi\leq\theta<\pi, 0\leq v<2\pi$ are the variables. It is obtained by gluing two M\"obius bands
along their edges. 
Then the map with typical critical Pareto set is given by $u=(\sqrt{x_1^2+x_2^2},x_3)$. 

$\Theta$ is a stratified set, consisting in a finite number of segments (codimension 1 strata) where rank$(u')=1$,
terminating at codimension 2 strata (isolated points) where rank$(u')=0$.

We will only consider $\ell=m$ which leads to the simplest topology.\\

\noindent $\bullet$ {\it Elementary Pareto sets}: the case of quadratic polynomials in 2-D.\\

As in the one-dimensional case, 
quadratic polynomials in ${\bf R}^\ell$ provide useful examples of maps with a typical critical Pareto set. For $m=\ell=2$ we may take
\begin{eqnarray*}
&u_1(x)={1\over2}(x_1^2+x_2^2), \ u_2(x)={1\over2}(x_1^2+x_1x_2+x_2^2), \quad \Theta=\{0\}\\
&u_1(x)=x_1x_2, \ u_2(x)={1\over2}(x_1^2-x_2^2), \quad \Theta=\{0\}\\
&u_1(x)={1\over2}(x_1^2+x_2^2), \ u_2(x)={1\over2}(x_1^2-x_2^2), \quad \Theta=\{x_1=0\}\\
\end{eqnarray*}
More generally, the critical Pareto set of two elliptic or two hyperbolic quadratic linearly independent polynomials reduces to the origin,
and to a line for a ``mixed'' pair. \\ 

\section{\bf Matrix valued Hamiltonian systems with the Pareto property}
\medskip
Since we are interested in 2-spinors, 
we work with $m=\ell=2$ and ${\cal H}(x,p)={\cal H}(p)$. \\

\noindent {\bf Definition 2}: Let ${\cal H}\in C^\infty({\bf R}^2)\otimes{\bf R}^2\otimes{\bf R}^2$ be a (real) Hermitian matrix.
We say that ${\cal H}$ has the {\it Pareto property} iff
there exists a smooth map (but with possibly degenerate derivative) 
$u:{\bf R}^2\to{\bf R}^2$, and a (regular) Hermitian matrix $C(p)$ such that, locally, and
up to conjugation with an elliptic factor of the form ${\cal H}(p)\mapsto{\cal A}(p)^*{\cal H}(p){\cal A}(p)$, we have:
$${\cal H}(p)=u'(p)C(p)(u'(p))^*$$

\noindent {\it Remark}: 
A hint on this definition is the following: let $p\in{\bf R}^2$ such that
$u'_1(p)=0$, then the ``pure classical state'' ${1\choose0}\in N(p)$, and similarly for 
${0\choose1}$ when $u'_2(p)=0$. So any ``classical state'' ${x\choose y}$ is a superposition of ``classical pure states''.\\

These decompositions are local and sometimes can be checked only in the sense of germs. If exact, we say that ${\cal H}$
is ``integrable in Pareto sense''. Only in 1-D problems, one can consider general Hamiltonians of the form $H(x,p)$. \\

\section{\bf Pareto property and the quadratic Hamiltonian of Elasticity Theory }
\medskip
Because Hamiltonians depending on $p$ alone are not structurally stable, few Hamiltonian
systems verify Pareto property. We have~:\\

\noindent {\bf Theorem 1}: For $A=1/2$, the quadratic Hamiltonian of Elasticity Theory \eqref{1} is integrable in Pareto sense:
with $u_1(\xi,\eta)={1\over2\sqrt2}(\xi^2-\eta^2)$, $u_2(\xi,\eta)={1\over\sqrt2}\xi\eta$, we have
${\cal H}(\xi,\eta)=u'(\xi,\eta)\diag(2,1)(u'(\xi,\eta))^*$
with $u'(u')^*={1\over2}(\xi^2+\eta^2)\Id$. 
Let $\sigma_2=\begin{pmatrix}0&i\\-i&0\end{pmatrix}$, we have the ``skew-diagonalization''
\begin{equation}\label{3}
{1\over2}\sigma_2({\cal H}(\xi,\eta)-\tau^2)\sigma_2^*=u'(\xi,\eta)
\bigl(\diag(\frac{1}{2},0)-(\xi^2+\eta^2)^{-1}\tau^2\Id\bigr)(u'(\xi,\eta))^*
\end{equation}
\medskip
From this we can construct by inverse Fourier transformation semi-classical spinors  
${}^t(\phi,\psi)(x,y;\tau)$ that verify near ``Helmholtz equation'' corresponding to \eqref{1}, $\tau^2$
standing for the energy parameter. We write $\widetilde\phi$ for the Fourier transform w.r.t the space variables.
There are two linearly independent solutions ${}^t(\phi_1,\psi_1)(x,y;\tau)$ of the equation 
$$\bigl(\diag(\frac{1}{2},0)-(\xi^2+\eta^2)^{-1}\tau^2\bigr){}^t(\widetilde\phi_1,\widetilde\psi_1)=0$$ 
given by
\begin{equation}\label{4}
\phi_1(x,y,\tau)=J_0(\sqrt{x^2+y^2}\frac{\sqrt2\tau}{h}), \quad 
\psi_1=J_0(\sqrt{x^2+y^2}\frac{\tau}{h})
\end{equation}
and $\phi,\psi$ are derived from \eqref{3} by convolution integrals. From this it is standard to deduce 
the spectral properties of ${\cal H}$.\\

\noindent $\bullet$ {\it Relative stability of the Pareto property}\\

\noindent {\bf Theorem 2}: Let 
$${\cal H}(\xi,\eta)=\begin{pmatrix}\xi^2+{1\over2}\eta^2&{1\over2}\xi\eta\\{1\over2}\xi\eta&{1\over2}\xi^2+\eta^2\end{pmatrix}+
{\cal O}((\xi,\eta)^3)$$
Then ${\cal H}(\xi,\eta)$ has the Pareto property (at least in the sense of germs at 0), with 
$C(\xi,\eta)=C_0+{\cal O}(\xi,\eta)$, $u(\xi,\eta)=u_0(\xi,\eta)+{\cal O}((\xi,\eta)^2)$. Moreover there is 
a skew-diagonalization of type \eqref{3}, and we can find a set of solutions to the ``Helmholtz equation'' as in \eqref{4}.\\ 

The proof goes as in Poincar\'e-Dulac theorem \cite{Ar2}: Namely we seek for a ``new'' $u$ of the form $v=u+f$,
with $f={\cal O}(|\xi,\eta|^3)$, and a ``new'' $C$ of the form $\diag(2,1)+\begin{pmatrix}a&b\\b&d\end{pmatrix}$.
The upper-left matrix element of $v'{\cal H}(v')^*$ is given by
$\sqrt2L_1f_1+(\xi^2+{\eta^2\over2})+{1\over2}(a\xi^2-2b\xi\eta+d\eta^2)$,
with $L_1=2x\frac{\partial}{\partial x}-y\frac{\partial}{\partial y}$, which has a resonance 2:1.
Its lower-right matrix element is given by 
$\sqrt2L_2f_2+(\eta^2+{\xi^2\over2})+{1\over2}(a\eta^2+2b\xi\eta+d\xi^2)$
with $L_2=2y\frac{\partial}{\partial x}+x\frac{\partial}{\partial y}$.
The off-diagonal terms involve the term $L_1f_2+L_2f_1$. It turns out that we can solve
(at least perturbatively) this system, the coefficients $a=a_1\xi+a_2\eta+\cdots$, $b=b_1\xi+b_2\eta+\cdots$, $d=d_1\xi+d_2\eta+\cdots$ 
being determined to fulfill the compatibility conditions.
The skew-diagonalization of type \eqref{3} follows from the fact that 
$v'(\xi,\eta)(v'(\xi,\eta))^*={1\over2}(\xi^2+\eta^2)^2(\Id+{\cal O}(|\xi,\eta|)$ is close to a multiple of $\Id$. 
We can still construct semi-classical solutions, and their asymptotics in $h$ for small $E$, 
obtained by varying the argument of the Bessel functions.\\

\section{\bf Conclusion}
\medskip
We have made an attempt to extend the notion of ``non-degenerate critical point'' for a scalar Hamiltonian
to the notion of ``Pareto critical set'' for a $2\times2$ Hamiltonian system. We focussed to the case 
where $\Theta$ reduces to a point. Our analysis applies to
the Hamiltonian ${\cal H}(p)$ quadratic in momenta arising in Elasticity theory,
for a particular value of the coefficients. 
It allows to account for
the spectral properties of ${\cal H}$
together with the semi-classical spectral asymptotics 
of (relative, i.e. depending again only on $p$) perturbations of ${\cal H}$ near $E=0$. These Hamiltonians present a codimension 3
singularity of order 4 at $\Theta=\{0\}$.
One of the limitations of this approach 
is due to the fact that the dynamics associated with an matrix valued Hamiltonian depending on $p$ alone is of the type of a gradient-flow dynamics,
while this fails to be the case for generic Hamiltonians. When the system has the time reversal property, we think that the
notion of Nash equilibrium [Sh] associated with more general types of dynamics provided a better alternative to Pareto equilibrium.\\

\noindent {\it Acknowledgements}: I thank Ilya Bogaevsky for interesting discussions. This work has been partially supported by
the grant PRC CNRS/RFBR 2017-2019 No.1556 ``Multi-dimensional semi-classical problems of Condensed Matter Physics and Quantum Dynamics''.\\

\end{document}